\documentstyle[10pt,epsfig,amssymb,amstex]{article}
\newcommand{\bee}{\begin{equation}}
\newcommand{\ee}{\end{equation}}
\newcommand{\beqn}{\begin{eqnarray}}
\newcommand{\eeqn}{\end{eqnarray}}

\newcommand{\kf}{{\bf k}}
\newcommand{\lf}{{\bf l}}
\newcommand{\ef}{{\bf e}}

\newcommand{\Pam}{I \!\! P}

\newcommand{\bs}{\boldsymbol}

\newcounter{savefig}

\begin{document}
\begin{titlepage}                                                              
\hfill
\hspace*{\fill}
\begin{minipage}[t]{4cm}
DESY--96--260\\
hep-ph/9612415\\
\end{minipage}
\vspace*{2.cm}                                                                 
\begin{center}                                                                 
\begin{LARGE}                                                                  
{\bf 
Production of Charm Quark Jets in \\
DIS Diffractive Dissociation
}\\
\end{LARGE}                                                                    
\vspace{2.5cm}                                                  
\begin{Large}
{ 
H.\ Lotter
}
\\
\end{Large}
\end{center}
\vspace{1.5cm}
\begin{center}
II.\ Institut f.\ Theoretische Physik, 
Universit\"at Hamburg, \\Luruper Chaussee 149, 
D-22761 Hamburg
\footnote
        {Supported by 
         Bundesministerium f\"ur Forschung und
         Technologie, Bonn, Germany under Contract 05\,6HH93P(5) and
         EEC Program "Human Capital and Mobility" through Network
         "Physics at High Energy Colliders" under Contract
         CHRX-CT93-0357 (DG12 COMA).}
\end{center}                                                   
\vspace*{2.cm}                          
\begin{quotation}                                                              
\noindent
We present a calculation of open charm quark production in diffractive 
deep inelastic electron-proton scattering in a perturbative QCD 
framework. 
The cross section is proportional to the 
square of the gluon density and explicitly displays breaking of 
Regge factorization. 
Jet cross sections as well as the charm contribution to 
the diffractive structure function are calculated.  
As a consequence of the steep rise of the gluon density at 
small $x$ the charm contribution to $F_2^{D}$ rises with decreasing
$x_{\Pam}$.
\end{quotation}                                                                
\vfill
\vspace{1cm}
\end{titlepage}
\begin{figure}[!h]
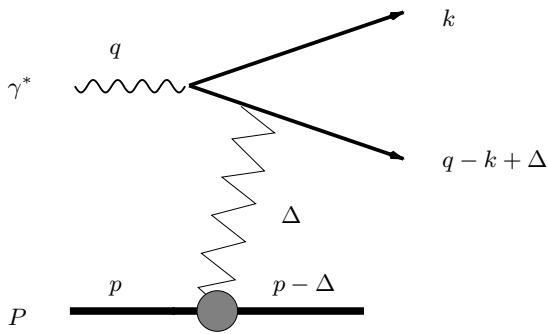

\begin{center}
\input diffqq1.pstex_t
\caption{ Diffractive production of a quark-antiquark pair 
in photon-proton scattering.
}
\label{fig1}
\end{center}
\end{figure}

\begin{figure}[!h]
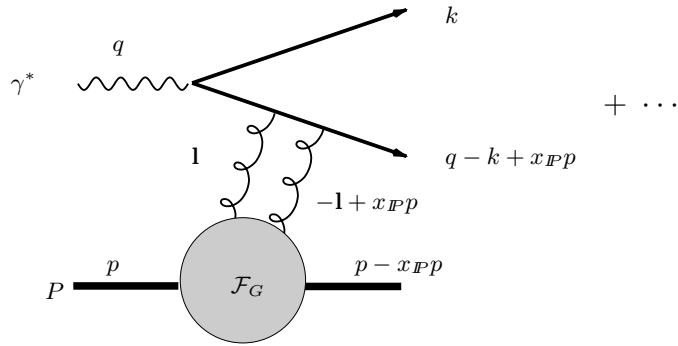

\begin{center}
\input diffqq2.pstex_t
\caption{ 
Representation of the amplitude in terms of the unintegrated gluon 
structure function. The dots represent three other diagrams which 
are generated by attaching the gluons to the quark lines in all 
possible ways. 
\label{fig2}
}
\end{center}
\end{figure}

                                                                
\noindent
{\bf 1.)} 
In the context of the discussion of rapidity gap events 
in deep inelastic electron-proton scattering 
observed at HERA a subclass 
of diffractive events in which a large mass scale appears in the diffractively
produced hadronic final state has created particular interest.
Representatives of this type of events are the 
diffractive production of heavy
vector mesons \cite{vecmes}, diffractive jet production 
\cite{blotwue,belotwue,diehl,niko,rysbec}
and diffractive production of 
open charm \cite{nikolaev,teubner}.   
Due to the presence of the large mass scale these processes offer the 
possibility to apply and test perturbative QCD in 
the setting of diffractive scattering.
The common feature of the above cited processes is the 
dependence of the cross section on the square of the gluon density of the 
proton. Because of this strong sensitivity these events have been
considered as a possible probe of the gluon density.
\\
In the above list the process of diffractive open charm production
is particularly promising. Compared to heavy vector meson production
it does not depend on a meson wave function which is poorly determined
from the theoretical side. Compared to jet production  
charm production does not require
a large transverse momentum which in turn leads to a strongly 
suppressed event rate.  
\\
In this letter we generalize our preceding calculations
on jet production \cite{blotwue,belotwue} in DIS diffractive
dissociation to finite quark mass. This allows, in particular, the 
investigation of open charm production. 
Our calculation is based on an analytical expression for the 
unintegrated gluon structure function which enables us to take into account 
a subset of subleading corrections which turn out to be numerically 
important \cite{blotwue}.
As the new contribution of the present work
we calculate jet cross sections for charm
quarks and compare with massless flavours.
In addition we compare the magnitude of the jet cross section
calculated in our model with the predictions of a model which is based
on nonperturbative two-gluon exchange \cite{diehl}.
Furthermore our formulae for the elecron-proton cross section include 
the dependence of the cross section on the angle of the jet plane relative
to the electron plane. 
For the case of the large charm
mass we can extend our expressions to low transverse momenta and can 
even integrate the transverse momentum to obtain the charm contribution 
of the diffractive structure function.     
We discuss the $x_{\Pam},\beta$ and $Q^2$ dependence of the
charm contribution to $F_2^{D(3)}$.
In this part of our analysis we obtain results similar to the ones 
presented in \cite{nikolaev} and \cite{teubner}.
\\
{\bf 2.)} 
The kinematics of the process is well-known and we only give the 
key ingredients. 
To calculate first the hadronic tensor $H_{\mu\nu}$ consider
the photon-proton subprocess
$\gamma^*(q) + P(p) \rightarrow c(k)+\bar{c}(q-k+\Delta)+P(p-\Delta)$.
We define the momenta of the particles as indicated in 
figure \ref{fig1}.  
The calculation is performed for zero momentum transfer $t=\Delta^2=0$.
For the momenta $k$ and $\Delta$ we use a Sudakov decomposition 
w.\ r.\ t.\ the light cone momenta $p$ and $q'=q+xp$ ($x=Q^2/(2pq)$)
\beqn
k&=&\alpha q'+ \beta p +\kf ,\\
\Delta&=& \alpha_{\Delta} q' + x_{\Pam}p + {\bs \Delta}.
\eeqn 
Using the mass shell conditions 
for the outgoing particles
one can show that we have 
$\alpha_{\Delta}={\cal O}({\bs \Delta^2}/(2pq'))$.
Assuming ${\bs \Delta}^2 \ll pq'$ and hence 
neglecting $\alpha_{\Delta}$ and ${\bs \Delta}$ we can cast the 
phase space in the form 
\beqn
d \Gamma = \frac{\pi}{8 pq}\frac{1}{\sqrt{1-4 \frac{\kf^2+m_c^2}{M^2}}}
dM^2 \,dt\,d\kf^2
\eeqn
with the charm quark mass $m_c^2$ and $M^2$ being the invariant mass 
of the $c\bar{c}$ pair which is related to the light cone momentum fraction
$\alpha$ through the relation 
\beqn
M^2=\frac{\kf^2+m_c^2}{\alpha(1-\alpha)}.
\eeqn
Energy-momentum conservation then leads to the phase space 
restriction $M^2 \geq 4(\kf^2+m_c^2)$.
Furthermore the longitudinal momentum fraction $x_{\Pam}$ transferred 
from the proton to the $c\bar{c}$ pair is fixed as 
\beqn
x_{\Pam}=\frac{M^2+Q^2}{W^2+Q^2}.
\eeqn
$W^2=(p+q)^2$ is the cms-energy of the photon-proton system. 
Another often used variable is $\beta$ defined as
\beqn
\beta=\frac{Q^2}{Q^2+M^2}
\eeqn
from which follows $\beta=x/x_{\Pam}$.
\\
Now we have to specify the coupling of the $c\bar{c}$ pair to the proton.
As the simplest model for an interaction in which no color is transferred
from the proton to the $c\bar{c}$ pair we take two-gluon exchange.
Since the charm quark is sufficiently heavy we treat both gluons 
perturbatively. We then make use of high-energy factorization \cite{catani} 
to express the amplitude of the photon-proton subprocess 
in terms of the unintegrated gluon density of the proton 
(fig.\ \ref{fig2})
\beqn
H_{\mu\nu}=\left(\int d^2 \lf \,C_{\mu}(\lf;\kf,Q^2,M^2)\, 
{\cal F}_G(x_{\Pam},\lf^2)\right)
\left(\int d^2 \lf' \,C_{\nu}(\lf';\kf,Q^2,M^2)\, 
{\cal F}_G(x_{\Pam},\lf{'}^2)\right)^*
.
\label{fac}
\eeqn
This factorization is valid in the leading-log($1/x_{\Pam}$) approximation
in which the imaginary part of the diagrams in fig.\ \ref{fig2}
contributes. In this approximation the difference of the 
longitudinal momenta of the two gluons is neglected. It is therefore
legitimate to use the same diagonal gluon density  
${\cal F}_G(x_{\Pam},\lf^2)$ which appears in inclusive DIS.
The coefficient
function
$C_{\mu}$
represents the upper part of the diagrams in fig.\ \ref{fig2}
with the gluons coupled to the quarks in all possible ways.
The calculation of the respective diagrams is straightforward 
if one uses the fact  
that the gluon polarization proportional to 
$p^{\sigma}$ gives the dominant contribution in the small-$x$ limit
\cite{c1,c2,c3}.
\\
In order to obtain the electron-proton cross section
we have to contract $H_{\mu\nu}$ with the usual lepton tensor.
This yields the following expression
\beqn
\frac{d \sigma^{e P}}
{d y d Q^2 d M^2 d \kf^2 d \phi d t}_{|t=0}
= \frac{\alpha_{\mbox{\tiny em}}}{2 y Q^2 \pi^2}
\left[
\frac{1+(1-y)^2}{2}
\frac{d \sigma^{\gamma^* P}_{T_1}}{d M^2 d \kf^2 d t}_{|t=0}
- 2 (1-y) \cos 2 \phi
\frac{d \sigma^{\gamma^* P}_{T_2}}{d M^2 d \kf^2 d t}_{|t=0}\right.
\nonumber \\ 
\left. + (1-y) \frac{d \sigma^{\gamma^* P}_{L}}{d M^2 d \kf^2 d t}_{|t=0}
+(2-y)\sqrt{1-y} \cos \phi
\frac{d \sigma^{\gamma^* P}_{I}}{d M^2 d \kf^2 d t}_{|t=0}
\right] \;\;\;.\;\;\;
\label{eq1}
\eeqn
The indices $T$ and $L$ refer to the contributions of transversely 
and longitudinally polarized photons. The term with index
$I$ is the interference contribution.
The angle $\phi$ is defined in the two-dimensional transverse plane
through the scalar product $\ef\cdot \kf_1 = |\ef| |\kf| \cos\phi$. 
Here $\ef$ is the transverse component of the electron 
momentum $e_{\mu}=1/y \,q'_{\mu}+x(1-y)/y \,p_{\mu}+\ef_{\mu}$
and $\kf_1$ is the transverse momentum of the 
charm (anti)quark which points into the proton hemisphere
(in the $c\bar{c}$ cm-system).
Note the negative sign of the $\cos 2 \phi$-term which is a characteristic
feature of the two-gluon exchange model. In photon-gluon fusion (one-gluon
exchange) this term comes with positive sign \cite{belotwue}.
\\
The expressions for the subprocess cross sections read
\beqn
\frac{d \sigma^{\gamma^* P}_{T_1}}{d M^2 d \kf^2 d t}_{|t=0} &=&
\frac{1}{M^4}\frac{1}{\kf^2}
\frac{1}{12}
e_c^2 \alpha_{\mbox{\tiny{em}}}\pi^2 \alpha_s^2
\frac{\kf^2+m_c^2}{\sqrt{1-4 \frac{\kf^2+m_c^2}{M^2}}}
\left[
\left(1-2 \frac{\kf^2+m_c^2}{M^2}\right)
\;[I_T(Q^2,M^2,\kf^2,m_c^2)]^2
\nonumber \right. \\
& &\left. + 
\;\; m_c^2 \;\; 
\frac{\kf^2 M^4}{(\kf^2+m_c^2)Q^4}
\;[I_L(Q^2,M^2,\kf^2,m_c^2)]^2
\right] 
\label{tra}
\\
\frac{d \sigma^{\gamma^* P}_L}{d M^2 d \kf^2 d t}_{|t=0} &=&
\frac{1}{M^4}\frac{1}{Q^2}
\frac{4}{3}
e_c^2 \alpha_{\mbox{\tiny{em}}}\pi^2 \alpha_s^2
\frac{\kf^2+m_c^2}{\sqrt{1-4 \frac{\kf^2+m_c^2}{M^2}}}
\;[I_L(Q^2,M^2,\kf^2,m_c^2)]^2
\label{long}
\\
\frac{d \sigma^{\gamma^* P}_I}{d M^2 d \kf^2 d t}_{|t=0} &=&
\frac{1}{M^4}\frac{1}{\sqrt{\kf^2Q^2}}
\frac{1}{3}
e_c^2 \alpha_{\mbox{\tiny{em}}}\pi^2 \alpha_s^2
(\kf^2\!+\!m_c^2) 
I_T(Q^2,M^2,\kf^2,m_c^2)\!I_L(Q^2,M^2,\kf^2,m_c^2)
\label{interf}
\\
\frac{d \sigma^{\gamma^* P}_{T_2}}{d M^2 d \kf^2 d t}_{|t=0}
&=&
\frac{1}{M^6}\frac{1}{\kf^2}
\frac{1}{12}
e_c^2 \alpha_{\mbox{\tiny{em}}}\pi^2 \alpha_s^2
\frac{(\kf^2+m_c^2)^2}{\sqrt{1-4 \frac{\kf^2+m_c^2}{M^2}}}
\; [I_T(Q^2,M^2,\kf^2,m_c^2)]^2
\label{asy}
\eeqn
The factor $e_c$ denotes the charge of the charm quark.
The essential dynamics is contained in the universal functions
$I_L,I_T$ for which we have the expressions
\beqn
I_L(Q^2,M^2,\kf^2,m_c^2) \!\!\!
&=& \!\!\!
-\int \frac{d \lf^2}{\lf^2}
{\cal F}_G(x_{\Pam},\lf^2)
\left[
\frac{Q^2}{M^2+Q^2}-\frac{(\kf^2+m_c^2)Q^2}{M^2 \sqrt{P_{\kf,\lf}}}
\right]
\label{il}
\\
I_T(Q^2,M^2,\kf^2,m_c^2) \!\!\!
&=& \!\!\!
-\int \frac{d \lf^2}{\lf^2}
{\cal F}_G(x_{\Pam},\lf^2)
\left[
\frac{2 M^2\kf^2}{(\kf^2+m_c^2)(Q^2+M^2)}-1
\right. \nonumber \\ 
& &\left.
            \phantom{xxxxx\frac{2 M^2\kf^2}{(\kf^2+m_c^2)(Q^2+M^2)}}
+\frac{\lf^2+\frac{\kf^2}{M^2}(Q^2-M^2)+m_c^2(1+\frac{Q^2}{M^2})}
{\sqrt{P_{\kf,\lf}}}
\right]
\label{it}
\eeqn
with $P_{\kf,\lf}$ being defined as
\beqn
P_{\kf,\lf}=(\lf^2+\frac{\kf^2}{M^2}(Q^2-M^2)
+\frac{m_f^2}{M^2}(Q^2+M^2))^2
+4 \kf^2( \frac{\kf^2}{M^2} Q^2+\frac{m_f^2}{M^2}(Q^2+M^2))
\eeqn
These expressions simplify considerably in the case of massless flavors 
($m_f^2=0$) \cite{blotwue,belotwue}.
To proceed one has to use an explicit model for the unintegrated
gluon structure function ${\cal F}_G$.
Within the leading-log($1/x_{\Pam}$) approximation it is consistent 
to express ${\cal F}_G$ as the solution of the BFKL equation \cite{bfkl}.
The $\lf$-integration in eqs.\ (\ref{il},\ref{it}) can then be performed 
analytically. The main outcome of this calculation is the determination 
of the relevant scale $\Delta^2$ of the process which is found to be 
\beqn
\Delta^2=(\kf^2+m_c^2)\frac{1}{1-\beta}
\eeqn
If we assume that this effective scale is large (compared to the 
inverse size of the proton) we can evaluate the integrals 
in eqs.\ (\ref{it},\ref{il}) in the leading-log($\Delta^2$) approximation. 
Using the relation 
\beqn
\int^{\Delta^2} d\lf^2 {\cal F}_G(x_{\Pam},\lf^2)  
=x_{\Pam}G(x_{\Pam},\Delta^2)
\eeqn
we can express the functions $I_L,I_T$ within this approximation in
terms of the gluon density $x_{\Pam}G(x_{\Pam},\Delta^2)$.
The result reads
\beqn
I_L
&=& 
\frac{(\kf^2+m_f^2)Q^2}{\kf^4 M^2}\frac{\xi-1}{(1+\xi)^3}
x_{\Pam}
G(x_{\Pam},\frac{\kf^2+m_c^2}{1-\beta}) ,
\label{ildla}
\\
I_T
&=&
\frac{4}{\kf^2}\frac{\xi}{(1+\xi)^3}
x_{\Pam}
G(x_{\Pam},\frac{\kf^2+m_c^2}{1-\beta}).
\label{itdla}
\eeqn
Here we have introduced, for brevity, the scaling variable 
$\xi=(\beta+m_c^2/\kf^2)/(1-\beta)$.
\\
Inserting these expressions into the cross section formulae we obtain the 
cross section for diffractive $c\bar{c}$ production in DIS in the 
double leading logarithmic approximation (DLA).
In \cite{blotwue} a certain class of corrections to the DLA was 
obtained in the massless case which turned 
out to be numerically important in the 
small-$\beta$ region. The same type of correction can be calculated in the 
massive case and the result for the correction terms reads
\beqn
I_L^{(c)}
&=&
\frac{(\kf^2+m_f^2)Q^2}{\kf^2M^2}
\frac{1}{(1+\xi)^3}\left(2+(1-\xi) \, \log\frac{\xi}{1+\xi}\right)
\frac{\partial}{\partial \kf^2}  x_{\Pam} G(x_{\Pam},\kf^2(1+\xi))
\\
I_T^{(c)}
&=&
2
\frac{1}{(1+\xi)^3}\left(1 - \xi -2\xi \log\frac{\xi}{1+\xi}\right)
\frac{\partial}{\partial \kf^2}  x_{\Pam} G(x_{\Pam},\kf^2(1+\xi))
\eeqn
These terms have to be added to the respective contributions 
(\ref{ildla}) and (\ref{itdla}). 
In the next section we turn to the numerical evaluation of our 
formulae.
\\
{\bf 3.)} 
As to the numerical evaluation we start with some preliminary remarks.
In all subsequent calculations we have integrated over the momentum transfer
$t$ after multiplication of the results of {\bf 2.)} with the elastic proton
form factor \cite{donnla}
\beqn
F_P(x_{\Pam},t) = 
\frac{4-2.8 \frac{t}{\mbox{{\tiny GeV}}^2}}{4 -
\frac{t}{\mbox{{\tiny GeV}}^2}}
\left(1-\frac{t}{0.7 \, \mbox{{\small GeV}}^2}\right)
x_{\Pam}^{-0.25 \frac{t}{\mbox{{\tiny GeV}}^2}}.
\eeqn 
For the charm quark mass we have taken the 
value $m_c=1.5 \,\mbox{GeV}$ and we have used the one-loop formula for 
$\alpha_s$ evaluated at the scale $(\kf^2+m_c^2)/(1-\beta)$.
In a preceding publication \cite{blotwue} we have discussed whether the 
leading order or the next-to-leading order gluon density should be 
used for a numerical prediction. We have presented some evidence that 
the NLO density is appropriate and consequently we perform all 
calculations with the GRV \cite{grv} NLO gluon density.
\\
First we present results for jet cross sections, i.\ e.\ 
we calculate the total electron proton cross section with a lower 
cutoff $\kf_0^2 \geq 2 \,\mbox{GeV}^2$ imposed on the transverse 
momentum integration. The other variables are integrated in a region 
which is adapted to the HERA kinematics, namely we choose the cuts
$Q^2 \geq 10 \,\mbox{GeV}^2$, $x_{\Pam} \leq 10^{-2}$ and 
$50 \, \mbox{GeV} < W < 220 \,\mbox{GeV}$. 
The results for the integrated cross section are listed in table \ref{tab1}.
\begin{table}[!h]
\begin{center}
\begin{tabular}{|c|c|c|c|}   \hline
 & $\kf^2_0 = 2 \,\mbox{GeV}^2$& $\kf^2_0 = 4 \,\mbox{GeV}^2$  
 & $\kf^2_0 = 8 \,\mbox{GeV}^2$ \\
 \hline \hline
 \multicolumn{4}{|c|} {DLA + corrections}  \\ \hline \hline
 $\sigma^{eP}_T$         & 27     & 11 & 3.4    \\
  \hline
 $\sigma^{eP}_L$        &  2.3   & 0.4 & 0.1   \\
  \hline \hline
 $\sum_{i=T,L}\sigma^{eP}_i$
                        &  29.3    & 11.4  & 3.5\\
\hline 
\end{tabular}
\end{center}
\caption{Results for total $eP$-cross sections (in pbarn) of 
diffractive dijet production for charm quarks.
\label{tab1}
}
\end{table}
\begin{table}[!h]
\begin{center}
\begin{tabular}{|c|c|c|c|}   \hline
 & $\kf^2_0 = 2 \,\mbox{GeV}^2$& $\kf^2_0 = 4 \,\mbox{GeV}^2$ & 
 $\kf^2_0 = 8 \,\mbox{GeV}^2$ \\
 \hline \hline
 \multicolumn{4}{|c|} {DLA + corrections}  \\ \hline \hline
 $\sigma^{eP}_T$    & 108     & 30 & 6\\
  \hline
 $\sigma^{eP}_L$    &  9   & 2 &  0.6\\
  \hline \hline
 $\sum_{i=T,L}\sigma^{eP}_i$
                    &  117    & 32  & 6.6\\
\hline 
\end{tabular}
\end{center}
\caption{Results for total $eP$-cross sections (in pbarn) of 
diffractive dijet production for three massless flavors. 
\label{tab2}
}
\end{table}
For comparison we quote in table \ref{tab2} also the 
results which are obtained
for three massless quarks \cite{blotwue}.
\\
It can be clearly seen that the relative contribution of the charm 
quarks compared to the massless flavors is dependent on the lower 
momentum cutoff $\kf_0^2$. The ratio 
$r=\sigma_{c\bar{c}}/\sum_{f=u,d,s}\sigma_{f\bar{f}}$ rises from
$r=0.25$ for $\kf_0^2=2\,\mbox{GeV}^2$ to 
$r=0.53$ for $\kf_0^2=8\,\mbox{GeV}^2$. 
This is easily understood since in the limit $m_c^2/\kf_0^2 \to 0$
the ratio $r$ should approach $2/3$ which follows from counting 
the electromagnetic charges.
It is also noticeable that the decrease of the transverse cross section
with increasing $\kf_0^2$ is weaker in the case of charm production compared 
to the massless case. Stated differently,
the $\kf^2$-spectrum of the transverse 
contribution is flatter in the massive case. As a consequence 
the ratio of the longitudinal to the transverse contribution
is decreasing with increasing $\kf_0^2$ for charm production whereas 
for the massless flavors it is approximately constant.
\\
The absolute numbers have been compared to the ones computed in a model 
based on nonperturbative two-gluon exchange (soft pomeron model) \cite{diehl}
already in \cite{procc}.
The soft pomeron model predicts cross sections 
which are reduced by a factor of 
approximately $1/2$ compared to our calculation.
This difference is mainly due to the rise of the cross section at small
$x_{\Pam}$ in our calculation according to the rise 
of the function $|x_{\Pam}G(x_{\Pam},\Delta^2)|^2$.
The soft pomeron model does not contain such a pronounced rise 
at small $x_{\Pam}$.
\\
So far we have discussed jet cross sections. Since the 
large charm quark mass
$m_c$ provides a hard scale $\Delta^2$ independent of the value of the 
transverse momentum $\kf^2$ we can also extend the perturbative calculation
to low values of $\kf^2$.   
We can even integrate over the transverse momentum which gives 
us the possibility to calculate the charm contribution to the 
diffractive structure function $F_2^{D(3)}$ in our model.
The latter has been introduced to describe diffractive events
(events with a rapidity gap) in DIS. In terms of 
$F_2^{D(3)}$ the diffractive 
electron proton cross section reads
\beqn
\frac{d \sigma_{\mbox{\tiny{DIFF}}}^{eP}}{d \beta d Q^2 d x_{\Pam}}
= \frac{2 \pi \alpha_{\mbox{\tiny{em}}}^2}{\beta Q^4}
[1+(1-y)^2]\; F_2^{D(3)}(\beta,Q^2,x_{\Pam})
\eeqn 
where the longitudinal contribution has been neglected. 
With this definition of $d \sigma_{\mbox{\tiny{DIFF}}}^{eP}$, 
$F_2^{D(3)}$ can be obtained 
from the photon-proton cross section
in the following way
\beqn
F_2^{D(3)}(\beta,Q^2,x_{\Pam}) = 
\frac{Q^2}{4 \pi^2 \alpha_{em}}\int_0^{\infty}d t
\int_0^{M^2/4-m_c^2}
d \kf^2 
\left[
\frac{d \sigma_{T_1}^{\gamma^*P}}{d x_{\Pam}d \kf^2 d t}
+ 
\frac{d \sigma_L^{\gamma^*P}}{d x_{\Pam}d \kf^2 d t}
\right].
\eeqn 
where the cross sections in brackets are obtained from multiplying
the expression for $t=0$ in eqs.\ (\ref{tra}) and (\ref{long}) 
with the proton form factor 
$F_P(x_{\Pam},t)$ and the factor $2 \pi$ (integration of $\phi$). 
\begin{figure}[!h]
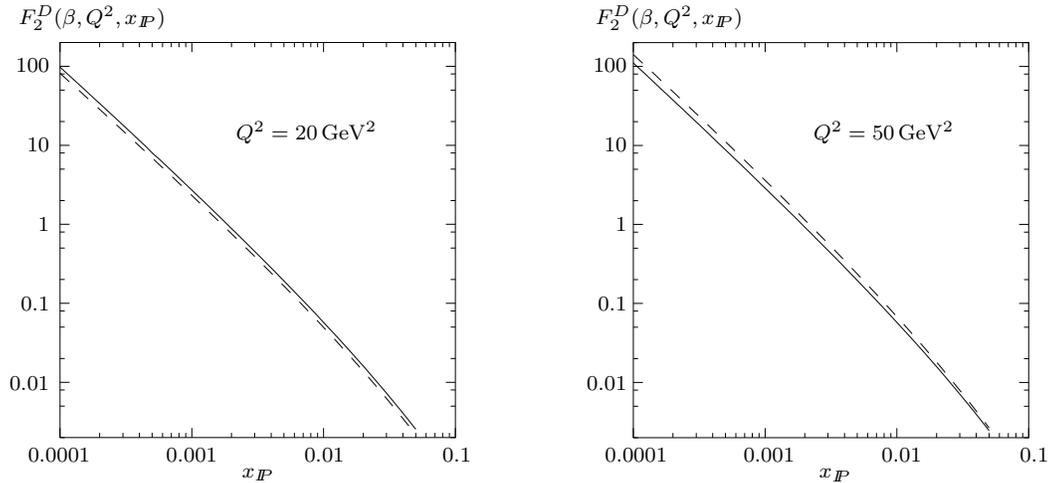

\begin{center}
\input xspec.pstex_t
\end{center}
\caption{
The $x_{\Pam}$-dependence of the diffractive structure function for 
$Q^2=20\,\mbox{GeV}^2$, $Q^2=50 \,\mbox{GeV}^2$ and
$\beta=1/3$, (solid line)
and $\beta=2/3$ (dashed line).
\label{fign1}
}
\end{figure}
In fig.\ \ref{fign2} we display $F_2^{D(3)}$ as a function of 
$x_{\Pam}$. First of all, these curves demonstrate the rapid increase
of the diffractive structure function at low values of $x_{\Pam}$.
The fact that the shapes of the curves are not identical for different 
$\beta$ indicates the (weak) breaking of Regge factorization.
Variation of $\beta$ leads to a 
variation of the $x_{\Pam}$-dependence. More precisely, a lowering 
of $\beta$ leads to an increase of the slope of the curve in accordance 
with the fact that the relevant scale of the gluon density 
is proportional to $1/(1-\beta)$. For smaller $Q^2$ (left hand side)
this effect is very small.
\begin{figure}[!t]
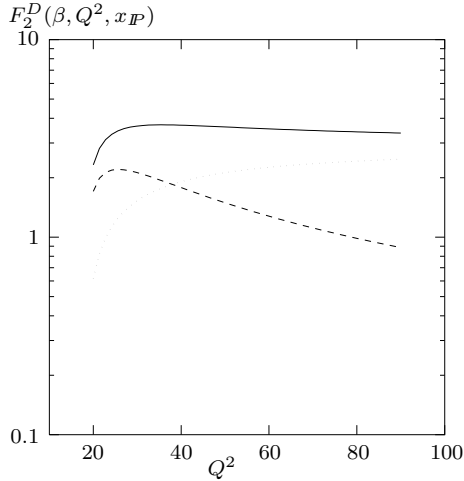

\begin{center}
\input qspec.pstex_t
\end{center}
\caption{
The $Q^2$-dependence of the diffractive structure function 
for $\beta=2/3$ and
$x_{\Pam}=10^{-3}$. The dotted line is the transverse contribution 
the dashed line is the longitudinal part and the solid lines 
represent the sum.
\label{fign2}
}
\end{figure}
By comparing the results for $Q^2=20 \,\mbox{GeV}^2$ and 
$Q^2=50 \,\mbox{GeV}^2$ one observes that a variation of $Q^2$ 
does not lead to a significant change of the absolute normalization
of $F_2^{D(3)}$. This is demonstrated explicitly in fig.\ \ref{fign2}
where the $Q^2$-dependence of $F_2^{D(3)}$, respectively its transverse and
longitudinal part, is displayed.
\begin{figure}[!h]
\begin{center}
\input beta.pstex_t
\end{center}
\caption{
The $\beta$-dependence of the diffractive structure function 
for $Q^2 = 20\,\mbox{GeV}^2$ and $Q^2 = 50\,\mbox{GeV}^2$ and 
$x_{\Pam}=10^{-3}$. The dotted line is the transverse contribution 
the dashed line is the longitudinal part and the solid line 
represents the sum.
\label{fign3}
}
\end{figure}
One observes a weak logarithmic increase of the transverse contribution
whereas the longitudinal part decreases rapidly with $Q^2$.
The longitudinal part, in other words, is a higher twist contribution.
\\
In fig.\ \ref{fign3} we show $\beta$-spectra for two different values of 
$Q^2$. The most noticeable feature here is of course the charm threshold
at the right end of each curve. One sees that near this threshold 
(for large $\beta$) the longitudinal contribution becomes comparable to
or even larger than the transverse contribution.
In the limit $\beta \to 0$ both the transverse and the longitudinal part 
tend to zero.
\\
We can compare our results with experimental data on inclusive diffractive 
DIS to roughly estimate the relative contribution 
of the $c\bar{c}$-production. The ZEUS collaboration gives a value 
of 30-40 for $F_2^{D(3)}(\beta,Q^2,x_{\Pam})$ at $x_{\Pam}=10^{-3}$,
$Q^2=16 \,\mbox{GeV}^2$ and $\beta=0.65$ \cite{zeus}. 
From the right hand side of fig.\ \ref{fign3} we find a value
of $F_2^{D(3)} \simeq 3.25$ for $x_{\Pam}=10^{-3}$,
$Q^2=20 \,\mbox{GeV}^2$ and $\beta \simeq 0.65$. We conclude that 
approximately $10\%$ of the diffractive events 
observed at HERA in the corresponding kinematical region 
are due to $c\bar{c}$-production. A similar estimate has been given in 
\cite{nikolaev}. One should emphasize that the relative contribution 
of $c\bar{c}$ is a function of $x_{\Pam}$ due to the strong rise of the 
$c\bar{c}$-cross section at low $x_{\Pam}$. Since the 
cross section of the diffractive production of massless flavors is not 
expected to increase equally fast at low $x_{\Pam}$
due to the dominance of low momentum scales, the $c\bar{c}$ 
contribution is enhanced at small $x_{\Pam}$.    
The relative contribution of $10\%$ seems to be large enough to expect the
charm threshold to be observable in the experimenta data.
\\
{\bf 4.)} 
To summarize, we have calculated the cross section for the diffractive
production of an open 
$c\bar{c}$-pair in DIS. We have assumed two-gluon exchange
and have treated both gluons perturbatively due to the large charm 
quark mass. In the double logarithmic approximation the cross section 
is proportional to the squared gluon density 
$|x_{\Pam}G(x_{\Pam},\Delta^2)|^2$ of the proton with the scale 
$\Delta^2 =(\kf^2+m_c^2)/(1-\beta)$. We have first calculated 
jet cross sections where $\kf^2$ is kept large and found that the 
charm contribution rises relative to massless flavours with increasing 
$\kf^2$. For very high $\kf^2$ the relative charm contribution follows
from the electromagnetic charge factors. Since $m_c^2$ is sufficiently large
the transverse momentum $\kf^2$ can be integrated and the charm contribution
to the diffractive structure function can be calculated in our model. 
As a function of $x_{\Pam}$ the diffractive structure function
rises steeply at small $x_{\Pam}$ due to the rise of the gluon density.
Since $\beta$ enters explicitly the scale of the latter our 
formulae display the breaking of Regge factorization. As a function
of $Q^2$ the charm contribution to the diffractive structure function
is approximately constant, i.\ e.\ it displays a leading-twist behavior.
The most remarkable property of the $\beta$-spectrum is the 
charm threshold at large $\beta$.  
At small $x_{\Pam}$ where the charm contribution is relatively 
largest this threshold should be observable in the data. At small $\beta$
both the transverse and the longitudinbal cross section tend to zero.
\\
As far as $c\bar{c}$-production is concerned our calculation is complete
in the given approximation. The data on diffractive dissociation, 
however, show a constant behavior in the limit $\beta \to 0$.
This shows that our model is not adequate to describe diffractive 
dissociation in the complete $\beta$-range. As an important correction
the production of additional gluons has to be considered.
Taking one additional gluon into account, one obtains a constant 
cross section at $\beta=0$. Corrections of this type have been 
investigated in \cite{c3,mark-bar}. Their impact on diffractive charm 
production has been analyzed in \cite{teubner}.    
The validity of the cited calculations has been limited to a restricted 
kinematical range. A complete calculation of the relevant corrections 
remains still an important task for future work. 
\\ \\
{\bf Acknowledgments:} 
I have benefited very much from discussions with 
Prof.\ J.\ Bartels,
Dr.\ M.\ W\"usthoff, Dr.\ M.\ Diehl and C.\ Ewerz.
I am grateful to Prof.\ J.\ Bartels and  
C.\ Ewerz for carefully reading the manuscript.
The support from the Studienstiftung des Deutschen Volkes is gratefully
acknowledged.

\end{document}